# Control of Spin in La(Mn,Zn)AsO Alloy by Carrier Doping

*Xingxing Li,*[†] *Xiaojun Wu,*[†,§] *and Jinlong Yang*[†,*]

[†]Hefei National Laboratory of Physical Science at the Microscale, University of Science and Technology of China, Hefei, Anhui 230026, China.

[§]CAS Key Laboratory of Materials for Energy Conversion and Department of Materials Science and Engineering, University of Science and Technology of China, Hefei, Anhui 230026, China.

*To whom correspondence should be addressed:
Email: jlyang@ustc.edu.cn

**Abstract**

The control of spin without magnetic field is one of challenges in developing spintronic devices. In an attempt to solve this problem, we proposed a novel hypothetic La(Mn$_{0.5}$Zn$_{0.5}$)AsO alloy from two experimentally synthesized rare earth element transition metal arsenide oxides, *i.e.* LaMnAsO and LaZnAsO. On the basis of the first-principles calculations with strong-correlated correction, we found that the La(Mn$_{0.5}$Zn$_{0.5}$)AsO alloy is an antiferromagnetic semiconductor at ground state, while bipolar magnetic semiconductor at ferromagnetic state. Both electron and hole doping in the La(Mn$_{0.5}$Zn$_{0.5}$)AsO alloy induces the transition from antiferromagnetic to ferromagnetic, as well as semiconductor to half metal. In particular, the spin-polarization direction is switchable depending on the doped carrier's type. As carrier doping can be realized easily in experiment by applying a gate voltage, the La(Mn$_{0.5}$Zn$_{0.5}$)AsO alloy stands for a promising spintronic material to generate and control the spin-polarized carriers with electric field.

**TOC Graphic**

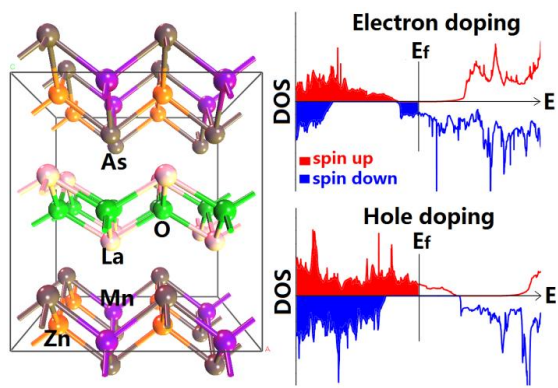

The control of spin of electrons in materials makes the basis of spintronics, which requires effectively generating, delivering, and manipulating the spin-polarized carriers.[1] In the past decades, various materials have been proposed for spintronic applications, including diluted magnetic semiconductors,[2] half metals,[3] spin gapless semiconductors,[4] and topological insulators[5] *etc*. For instance, half metal, which has one conducting spin channel and the other semiconducting one, is regarded as an ideal spintronic material to provide 100% spin-polarized carriers.[3] In developing a spintronic device, however, generating and manipulating the spin-polarized carriers without magnetic field is highly desirable, but facing many challenges.

In recent years, both experimental and theoretical works have proposed some schemes to realize the electrical control of spin-polarized carriers in many materials. For example, the electrical control of spin orientation can be achieved experimentally in a semiconducting structure with spatially varying *g*-tensor,[6] which provides a magnetic field gradient, or simply *via* spin-orbital coupling presented in most semiconducting materials.[7] Furthermore, a conceptual material of bipolar magnetic semiconductor with designed electronic band structures has been proposed recently for this purpose.[8] With the conduction band and valence band edges possessing inverse spin-polarization, bipolar magnetic semiconductors can be converted into half metals with controllable spin-polarization direction by applying an external electric field. Theoretically, some nanostructures and bulk materials have been predicted to be bipolar magnetic semiconductors.[9-11] So far, exploring new spintronic materials is still a long term task to achieve the electrical control of spin in spintronics materials.

The ZrCuSiAs-type compounds have recently been largely explored partly for the discovery of high-temperature superconductivity, such as the known "1111" family of iron-based superconductors.[12] Other interesting properties, including magnetic ordering[13] and transparent semiconducting behavior,[14] have also been studied for a long time. The ZrCuSiAs-type compounds family possesses many members for the combination of four different elements. Among them, the rare earth element transition metal arsenide oxides, LaMnAsO[15] and LaZnAsO,[16] are two typical examples with similar lattice constants (space group P4/nmm), which show diverse physical properties. For instance, LaMnAsO is an antiferromagnetic (AFM) semiconductor with pretty high Néel temperature of 317 K.[17] LaZnAsO is also a semiconductor and expected to be diluted magnetic semiconductor if $Zn^{2+}$ is partially replaced with other divalent transition metals.[18]

In this work, we proposed a new theoretical design of ZrCuSiAs-type La(Mn$_{0.5}$Zn$_{0.5}$)AsO structure, which is a hypothetic alloy of LaMnAsO[15] and LaZnAsO.[16] We demonstrated that the control of spin-polarization direction without magnetic field can be realized in the La(Mn$_{0.5}$Zn$_{0.5}$)AsO alloy. Our first-principles calculations with strong-correlated correction indicated that the La(Mn$_{0.5}$Zn$_{0.5}$)AsO alloy transits from AFM semiconductor to ferromagnetic (FM) half metal *via* carrier doping. In particular, the spin-polarization direction is switchable depending on the doped carrier's type, *i.e.* hole or electron.

As shown in Figure 1, the structure of the La(Mn$_{0.5}$Zn$_{0.5}$)AsO alloy can be looked as a mixture of LaMnAsO and LaZnAsO with 1:1 ratio. The lattice mismatch between LaMnAsO (a=b=4.124Å, c=9.030Å)[15] and LaZnAsO (a=b=4.095Å, c=9.068Å)[16] is smaller than 1%, implying an ideal mutual solubility. Actually, the optimized lattice constants of the used supercell containing 2×2×1 unit cells for the La(Mn$_{0.5}$Zn$_{0.5}$)AsO alloy (a=b=8.212Å and c=9.101Å) indicate that there is no obviously structural distortion. The La(Mn$_{0.5}$Zn$_{0.5}$)AsO alloy has a layered structure, where the [LaO]$^+$ and [(Mn,Zn)As]$^-$ layers with tetrahedral coordination patterns are stacked in an ABA sequence. The intralayer bonds are covalent, whereas the interlayer interaction is of an ionic type. Within the [(Mn,Zn)As]$^-$ layer, the Mn and Zn atoms are uniformly distributed. Test calculations showed that the clustering of Zn and Mn atoms in the [(Mn,Zn)As]$^-$ layer leads an energy increase of about 0.28 eV per supercell. Recent experimental progress suggests that the preparation of this alloy would be facile using the salt flux technique followed by an additional annealing procedure.[19]

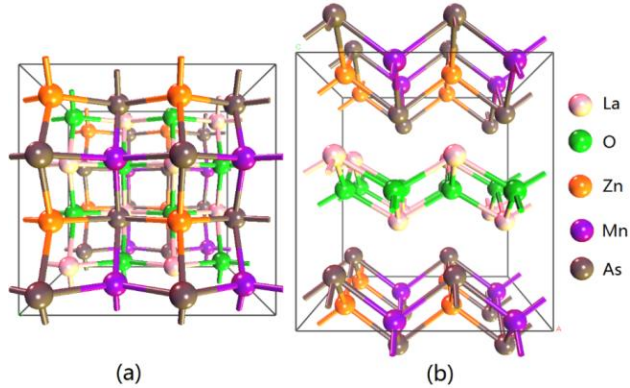

Figure 1. The (a) top view and (b) side view of La(Mn$_{0.5}$Zn$_{0.5}$)AsO crystal.

The electronic structure calculation indicates that the La(Mn$_{0.5}$Zn$_{0.5}$)AsO alloy prefers AFM magnetic coupling at its ground state. The FM state is less favorable than the AFM state with an energy difference of about 0.62 eV per supercell. The spin charge density distributions of both AFM and FM states are plotted in Figure 2a. It is clear that the magnetic moment of La(Mn$_{0.5}$Zn$_{0.5}$)AsO mainly locates on Mn atoms. The local magnetic moment of Mn atom is about 4.19 $\mu_B$, implying that Mn possesses about +2 $e$ charge with a high spin state. Note that the magnetic coupling between two neighboring [(Mn,Zn)As]$^-$ layers was not examined as the interlayer distance between them is about 9.1Å. Thus, the interlayer magnetic coupling is very weak and was set to be ferromagnetic based on the previous experiments.[17,20] The total density of states (DOS) of the La(Mn$_{0.5}$Zn$_{0.5}$)AsO alloy are plotted in Figure 2 for both AFM and FM states.

Clearly, the AFM La(Mn$_{0.5}$Zn$_{0.5}$)AsO alloy is an intrinsic semiconductor with a band gap of about 1.28 eV (Figure 2b), while its FM state exhibits a distinct band structure character of ideal bipolar magnetic semiconductor,[8] *i.e.* the valence band and conduction band approach the Fermi level with opposite spin polarization (Figure 2c). Therefore, it provides a possibility to tune the spin-polarization direction with electric field in the La(Mn$_{0.5}$Zn$_{0.5}$)AsO alloy at its FM state.

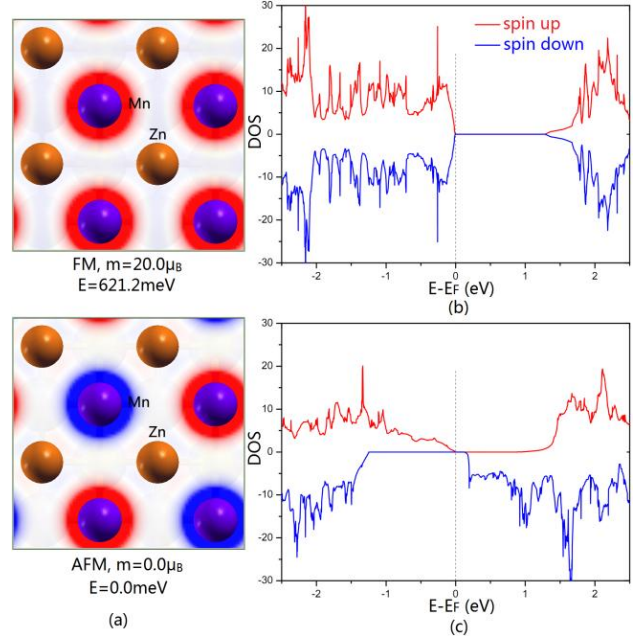

Figure 2. (a) The distribution of spin charge density in [(Mn,Zn)As]$^-$ layer under FM (upper part) and AFM (lower part) states. The isosurface value is 0.08e/Å$^3$. Red and blue indicate the positive (spin up) and negative (spin down) values, respectively. The calculated total DOS for La(Mn$_{0.5}$Zn$_{0.5}$)AsO under (b) ground AFM state and (c) FM state. The Fermi level is set to zero.

It is well known that the AFM-FM transition can be realized by various chemical and physical methods in many materials, such as strain[21] and carrier doping.[22-25] Here, we found that the La(Mn$_{0.5}$Zn$_{0.5}$)AsO alloy doped with few amount of carriers retains its AFM ground state. The magnitude of the energy difference between the AFM and FM states, however, decreases with the increasing doping concentration (yellow region in Figure 3a). Meanwhile, the carrier doping induces a semiconductor-metal transition, confirmed with the calculated DOS in Figure 3b and 3c. These results are not surprising since the profile of total DOS doesn't change much and the Fermi level just moves into either conduction or valence band when introducing a small amount of electrons or holes in the system.

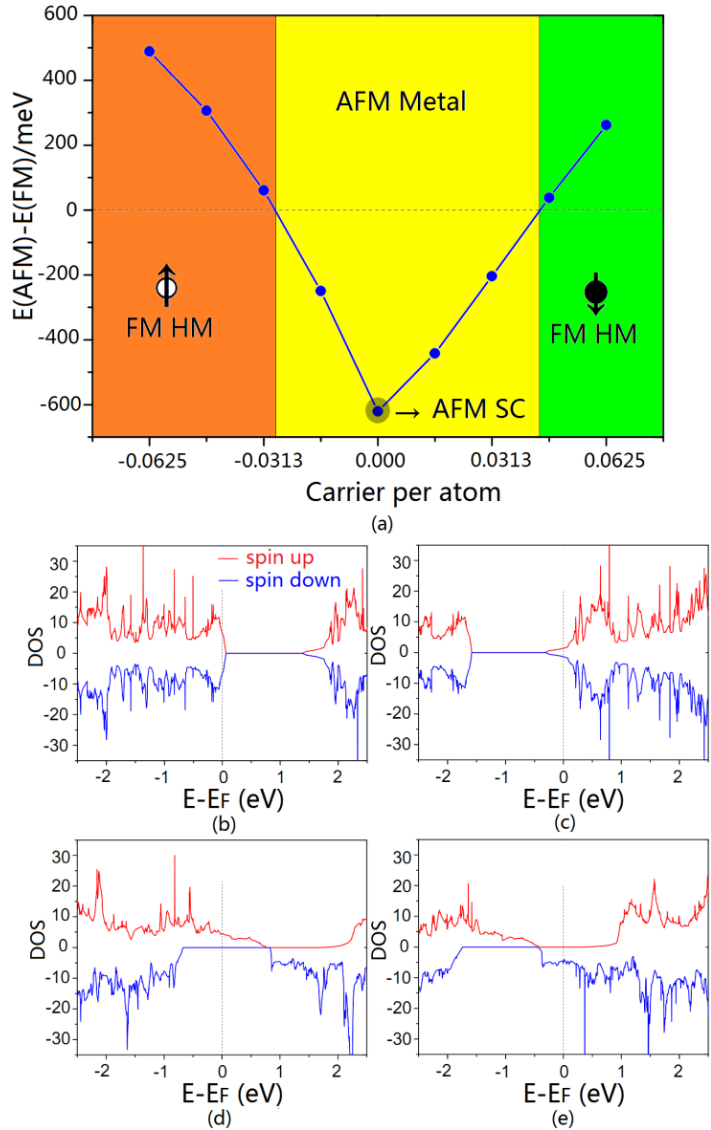

Figure 3. (a) The energy differences per supercell between AFM and FM states are plotted versus the concentration of doped carriers. The positive and negative values are for electron and hole doping, respectively. The up and down arrows indicate spin up and spin down, respectively.

Half metal and semiconductor are denoted as HM and SC, respectively. The total DOS of La(Mn$_{0.5}$Zn$_{0.5}$)AsO with the doping concentration of (b) 0.0156 holes/atom, (c) 0.0156 electrons/atom, (d) 0.0625 holes/atom and (e) 0.0625 electrons/atom. The Fermi levels are all set to zero.

Continually increasing the doping concentration induces a AFM-FM transition in the La(Mn$_{0.5}$Zn$_{0.5}$)AsO alloy, as shown in Figure 3a. This kind of transition has been observed experimentally in a doped LaMnAsO material.[26] Moreover, it can be concluded that the hole doping is more efficient than the electron doping to induce this transition, where the former has a lower concentration at the transition point than the latter. Since the valence and conduction bands near the Fermi level in FM La(Mn$_{0.5}$Zn$_{0.5}$)AsO are inversely spin polarized, introducing considerable electrons or holes in FM La(Mn$_{0.5}$Zn$_{0.5}$)AsO definitely turns it into a half-metallic material. In particular, the half-metallic La(Mn$_{0.5}$Zn$_{0.5}$)AsO alloy shows inverse spin-polarization, depending on the doped carrier's type. For instance, the hole-doped La(Mn$_{0.5}$Zn$_{0.5}$)AsO alloy would be fully spin-up polarized, while the electron-doped one is spin-down polarized, as shown in Figure 3d and 3e. This behavior presents a potential to alter the spin-polarization direction in La(Mn$_{0.5}$Zn$_{0.5}$)AsO by carrier doping, which can be realized in experiment just by altering the sign of applied gate voltage.

In essence, the electrical control of spin in the La(Mn$_{0.5}$Zn$_{0.5}$)AsO alloy is similar to the previously proposed bipolar magnetic semiconductors, where the spin direction of carriers is manipulated by moving the Fermi level to either conduction band or valence band under electric field.[8] The only difference is that the bipolar spin polarization in the La(Mn$_{0.5}$Zn$_{0.5}$)AsO alloy is obtained *via* carrier-doping induced AFM-FM phase transition. As shown in Figure 3d and 3e, the La(Mn$_{0.5}$Zn$_{0.5}$)AsO alloy shows a nearly vanished spin-flip gap between valence band and conduction band edges, and the big spin-conserved gaps are about 0.83 and 1.49 eV for spin-up and spin-down channels, respectively. Thus, the spin-polarization direction can be switched easily by a small shift of the Fermi level and the applied gate voltage can be tuned in a wide range while maintaining half-metallic conduction.

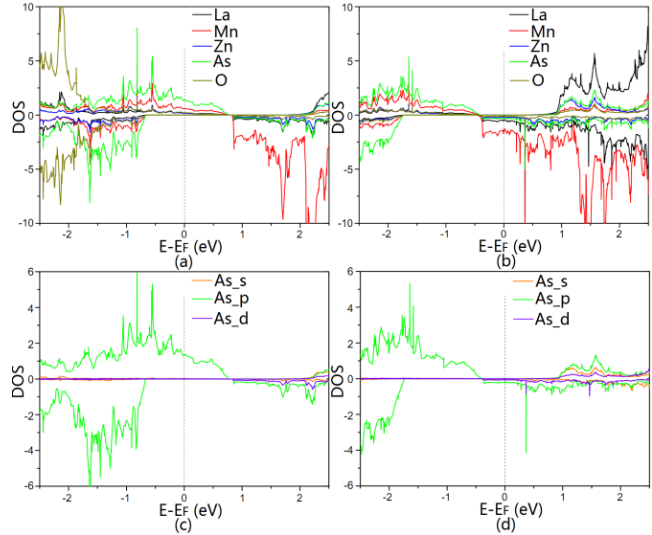

Figure 4. (a) and (b) are the site-projected DOS of La(Mn$_{0.5}$Zn$_{0.5}$)AsO with the doping concentration of 0.0625 holes/atom and 0.0625 electrons/atom, respectively. (c) and (d) are the orbital-projected DOS for As atoms under the same doping concentration. The Fermi energy levels are set to zero.

To understand the profound AFM-FM transition in the La(Mn$_{0.5}$Zn$_{0.5}$)AsO alloy, we plotted both the site- and orbital-projected DOS at the doping concentration of 0.0625 carriers per atom in Figure 4. In its neutral state, *i.e.* without doping, the strong superexchange interaction[27] mediated by nonmagnetic As atoms leads to the AFM coupling between adjacent Mn atoms. With high hole-doping concentration, La(Mn$_{0.5}$Zn$_{0.5}$)AsO exhibits a robust FM order and the states around Fermi level are mainly contributed by As' *4p* and Mn's *3d* orbitals (Figure 4a and 4c). The strong hybridization between them suggests that the ferromagnetism in the doped La(Mn$_{0.5}$Zn$_{0.5}$)AsO alloy originates from the Zener's *p-d* exchange interaction.[28,29] Usually, the Zener's *p-d* exchange interaction leads to a higher Curie temperature with a higher doping concentration, consistent well with the predicted increase of energy difference between AFM and FM states in the La(Mn$_{0.5}$Zn$_{0.5}$)AsO alloy (Figure 3a). Therefore, the magnetic order transition in the La(Mn$_{0.5}$Zn$_{0.5}$)AsO alloy is a competition result between the superexchange and Zener's *p-d* exchange interactions. For electron doping, however, the hybridization between As' *4p* and Mn's *3d* is weak (Figure 4b and 4d), which partly explains why the electron doping is less efficient than hole doping for this transition in the La(Mn$_{0.5}$Zn$_{0.5}$)AsO alloy.

In conclusion, we presented a new spintronic material La(Mn$_{0.5}$Zn$_{0.5}$)AsO alloy obtained from two experimentally synthesized rare earth element transition metal arsenide oxides LaMnAsO and LaZnAsO. The carrier doping induces a transition from AFM semiconductor to FM half metal in it. In particular, the spin-polarization direction in La(Mn$_{0.5}$Zn$_{0.5}$)AsO is switchable, depending on the type of doped carriers. The La(Mn$_{0.5}$Zn$_{0.5}$)AsO alloy serves as an example of a new type of spintronic materials which can provide completely spin-polarized currents with tunable spin-polarization direction by applying an external gate voltage. It is expected that the various analogous ZrCuSiAs-type compounds will form a promising family of precursors for exploring similar materials to enable the electrical control of spin. Very recently, a two dimensional "1111" diluted magnetic semiconducting (La$_{1-x}$Ba$_x$)(Zn$_{1-x}$Mn$_x$)AsO in bulk form has been synthesized,[30] implying the great potential of this type of materials in spintronic applications.

**Computational Methods**

The first-principles calculations were performed by using the density functional theory (DFT) method within the Perdew-Burke-Ernzerholf (PBE) generalized gradient approximation (GGA),[31] implemented in Vienna *ab initio* Simulation Package (VASP).[32] The strong-correlated correction was considered with GGA+U method[33] to deal with the Mn's *3d* and La's *4f* electrons. The effective onsite Coulomb interaction parameter (U) and exchange interaction parameter (J) are set to be 4.0 and 1.0 eV for Mn's *3d* electrons, respectively. For La's *4f* electrons, these values are 7.5 and 1.0 eV, respectively. These values have been tested and used in previous experimental and theoretical works.[34-37] Our test calculations also showed that changing U's value (3.0 and 5.0 eV) for Mn provides very similar results. For Zn atoms, we didn't apply the strong-correlated correction as their *3d* orbitals are fully occupied and locate far away from the Fermi energy level. The projector augmented wave (PAW) potential[38] and the plane-wave cut-off energy of 400 eV are used. A supercell model containing 2×2×1 unit cells is used to explore their magnetic orders, which includes 32 atoms consisting of 8 La, 4 Mn, 4 Zn, 8 As, and 8 O atoms. A Monkhorst-Pack k-point mesh of 7×7×5 is used. Both the cell shape, cell volume and the positions of all atoms are relaxed without symmetry constraint until the force is less than 0.01 eV/Å. The criterion for the total energy is set as $1\times10^{-6}$ eV.

**Acknowledgment**

This work is partially supported by the National Key Basic Research Program (2011CB921404, 2012CB922001), by NSFC (21121003, 91021004, 20933006, 11004180, 51172223), by Strategic Priority Research Program of CAS (XDB01020300), and by USTCSCC, SCCAS, Tianjin, and Shanghai Supercomputer Centers.